\documentclass[11pt, a4paper,UKenglish,cleveref,thm-restate,numberwithinsect]{article}
\usepackage[margin=1.25in]{geometry}
\usepackage{libertine}
\usepackage[T1]{fontenc}
\usepackage[utf8]{inputenc}
\usepackage{amsfonts,amssymb,amsmath,amsthm}
\usepackage{calc}
\usepackage{tikz}
\usetikzlibrary{calc}
\usepackage[bottom,para,symbol]{footmisc}
\usepackage{hyperref}
\usepackage{thmtools}
\usepackage{thm-restate}

\usepackage{pdfpages}
\usepackage{epigraph}
\usepackage{lineno}

\usepackage{url}
\usepackage{hyperref}
\usepackage{enumitem}
\usepackage{xspace}
\usepackage{todonotes}
\usepackage{thmtools}
\usepackage{mathrsfs}  
\usepackage{textpos}
\usepackage[normalem]{ulem}
\usepackage[capitalize, nameinlink]{cleveref}

\newtheorem{theorem}{Theorem}[section]
\newtheorem{lemma}[theorem]{Lemma}

\newtheorem{corollary}[theorem]{Corollary}

\newtheorem{conjecture}[theorem]{Conjecture}
\theoremstyle{definition}

\usepackage[utf8]{inputenc}
\usepackage{amsfonts,amssymb,amsmath}
\usepackage{tikz}
\usetikzlibrary{calc}
\usepackage{xspace}
\usepackage{mathrsfs}
\usepackage[ruled,vlined]{algorithm2e}
\usepackage{eso-pic}


\newcommand{\algtt}[1]{\ifmmode\text{\texttt{#1}}\else\texttt{#1}\fi}
\newcommand{\Oh}{\mathcal{O}}

\newcommand{\Weight}[1]{\ensuremath{\lambda}}

\newcommand{\weight}[1]{\ensuremath{\lambda}}
\newcommand{\latency}[1]{\ensuremath{\ell}}

\newcommand{\pred}{\mathtt{parent}}
\newcommand{\ea}{\mathtt{ea}}
\newcommand{\ld}{\mathtt{ld}}
\newcommand{\inedges}{\mathtt{in}}
\newcommand{\outedges}{\mathtt{out}}
\newcommand{\prefld}{{\textrm{\textsc{Prefld}}}}

\newcommand{\arr}{\mathsf{arr}}
\newcommand{\key}{\mathsf{key}}

\renewcommand{\leq}{\leqslant}
\renewcommand{\geq}{\geqslant}
\renewcommand{\le}{\leqslant}
\renewcommand{\ge}{\geqslant}
\renewcommand{\mid}{~|~}
\newcommand{\eps}{\varepsilon}

\SetKwFunction{Init}{\textrm{\textsc{Init}}}
\SetKwFunction{AddArc}{\textrm{\textsc{AddArc}}}
\SetKwFunction{RemoveArc}{\textrm{\textsc{RemoveArc}}}
\SetKwFunction{Dist}{\textrm{\textsc{EA}}}
\SetKwFunction{Path}{\textrm{\textsc{EAPath}}}

\SetKwFunction{AddEdge}{\textrm{\textsc{AddEdge}}}
\SetKwFunction{RemoveEdge}{\textrm{\textsc{RemoveEdge}}}

\usepackage{todonotes}

\title{Partially Dynamic Algorithms for Single-Source Temporal Paths}

\author{Davide Bilò \thanks{Department of Information Engineering, Computer Science, and Mathematics, University of
L’Aquila, Italy, \texttt{davide.bilo@univaq.it}. Supported by the project \emph{SOS-TG: Spanners and Oracles for Static and Temporal
Graphs}, funded by Università degli Studi dell’Aquila under the Call for Proposals for Fundamental Research
and Early-Career Research Grants – Year 2026.}
\and Bartłomiej Bosek \thanks{Institute of Theoretical Computer Science, Faculty of Mathematics and Computer Science, Jagiellonian University, Krak\'ow, Poland. \texttt{bartlomiej.bosek@uj.edu.pl}}
\and Jadwiga Czyżewska \thanks{University of Warsaw, Poland. \texttt{j.czyzewska@mimuw.edu.pl}. Supported by Polish National Science Centre SONATA BIS-12 grant number 2022/46/E/ST6/00143.}
\and Evangelos Kipouridis \thanks{Max Planck Institute for Informatics, Saarland Informatics Campus, Germany. \texttt{kipouridis@mpi-inf.mpg.de}.}
\and Karol Węgrzycki \thanks{Max Planck Institute for Informatics, Saarland Informatics Campus, Germany. \texttt{kwegrzyc@mpi-inf.mpg.de}. Supported by the Deutsche Forschungsgemeinschaft (DFG, German Research Foundation) grant number 559177164.}
\and Anna Zych-Pawlewicz \thanks{University of Warsaw, Poland \texttt{anka@mimuw.edu.pl}. Supported by the project BOBR that received 
funding from the European 
Research Council (ERC) under the 
European Union’s Horizon 2020 
research and innovation 
programme, grant agreement 
No. 948057.
}
}

\begin{document}
\begin{titlepage}
\date{}
\maketitle

\begin{abstract}
We present efficient partially dynamic algorithms for maintaining earliest arrival times from a fixed source in temporal digraphs with $n$ vertices, possibly with parallel arcs, where each arc is associated with a time label and a latency. 

For the incremental scenario where updates are arc insertions, our algorithm has a constant query time and requires a total update time of $\Oh(n + m \log m)$, where $m$ is the total number of arcs added to the arc-less graph. For the decremental scenario where updates are arc removals, our algorithm has a constant query time and requires a total update time of $\Oh(n + m \log m)$, where $m$ is the number of arcs in the graph upon its initialization. We also consider earliest-arrival queries with an arbitrary departure-time threshold and develop an incremental algorithm with $O(n+\Delta_s m\log m)$ total update time, $O(\log m)$ query time, and $O(n+m)$ space, where $\Delta_s$ is the number of distinct departure times of arcs leaving the source.
All of our algorithms are also capable of reporting an earliest arrival path from the source to the queried vertex $v$ in constant time per edge once the earliest arrival time at $v$ is known.

We complement our positive results with conditional lower bounds based on the Online Boolean Matrix-Vector Multiplication conjecture (a.k.a. OMv conjecture). For both the incremental and the decremental scenarios, we show the non-existence of an algorithm with $\Oh(n^{1-\varepsilon})$ amortized update time and $\Oh(n^{2-\varepsilon})$ amortized query time, even if the algorithm is randomized and latencies on the arcs are uniform. For arbitrary departure-time queries, we also establish an OMv-based conditional lower bound showing that our incremental algorithm is optimal up to logarithmic factors on dense temporal graphs. Thus, efficient total running time is essentially the best outcome we can get. Our conditional lower bounds hold also for the case of undirected temporal graphs.    
\end{abstract}

\def\thepage{}
\thispagestyle{empty}
 \begin{textblock}{20}(-1.8, 1.6)
  \includegraphics[width=40px]{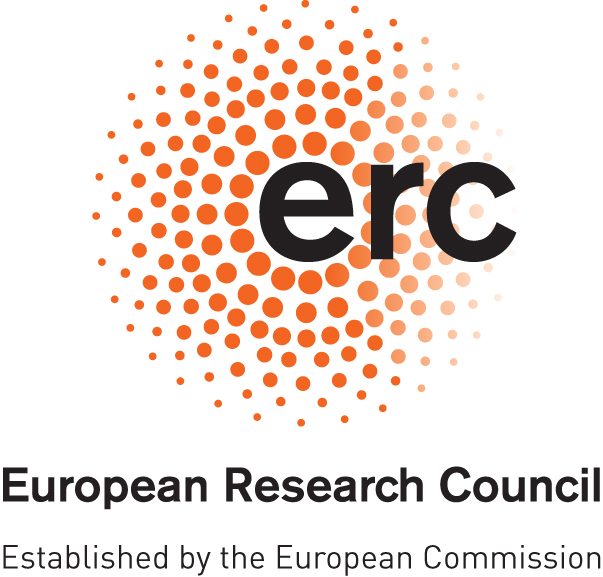}%
 \end{textblock}
 \begin{textblock}{20}(-2.45, 1.5)
  \includegraphics[width=90px]{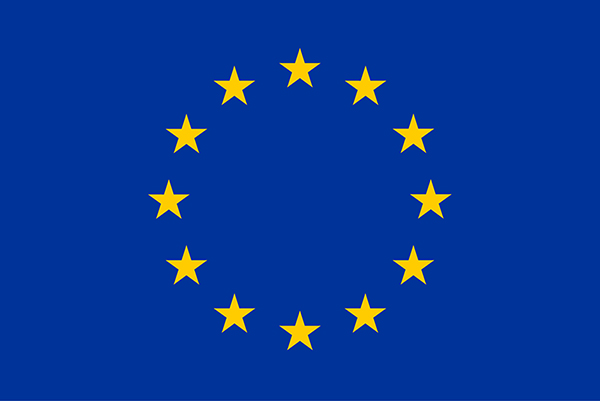}%
 \end{textblock}

\end{titlepage}

\setcounter{page}{1}
\hypersetup{pageanchor=true}
\section{Introduction}

\emph{Temporal graphs} are graphs in which edges are available only at specific time instants. They are a standard model for networks whose interactions change over time, such as communication networks, transportation systems, social networks, and contact networks. In these settings, the time at which an interaction occurs is often essential, and cannot be ignored without losing important information.

The most basic model of a temporal graph was introduced by Kempe, Kleinberg, and Kumar~\cite{KempeKK02}. In this model, a temporal graph consists of an underlying graph together with a set of integer labels assigned to each edge, representing the times at which the edge is available. As movement through the network must respect time, the basic notion is that of a \emph{temporal path}, i.e., a path whose edge labels appear in non-decreasing order. Several variants of this model have been studied, including strict temporal paths, paths with waiting constraints, continuous-time models, and stochastic models; see, e.g.,~\cite{CasteigtsFQS12, Holme14}.

Many algorithmic problems on temporal graphs have been considered in the literature. These include the computation of temporal paths optimizing different objectives, such as earliest arrival, minimum duration, or minimum number of hops~\cite{AkridaMNRSZ20,WuCHKLX14,CasteigtsHMZ21}; sparsification and temporal spanners~\cite{BiloDG0R22,BiloDGLR24,CasteigtsPS21,CasteigtsRRZ24,KempeKK02}; exploration problems~\cite{ErlebachKLSS19,ErlebachS23}; temporal versions of covering problems~\cite{AkridaMSZ20,HammKMS22}; and network creation games~\cite{BiloC0GKLS23, BiloC0GKLS25}. 
Among these topics, \emph{temporal reachability} is one of the most basic. Given two vertices $u$ and $v$, one asks whether there exists a temporal path from $u$ to $v$. Unlike the static case, the answer may depend on the starting time, the allowed waiting time, and the chosen path model. Moreover, temporal reachability is in general not symmetric, and its behavior is often more complex than standard graph connectivity.

Dynamic maintenance of reachability information is well studied for static graphs (see, e.g.,~\cite{HanauerHS22,HuangHKPT23} and the references therein). In contrast, the temporal case has only recently received attention. Brito, Albertini, Casteigts, and Travençolo~\cite{Brito2021ADD} gave a data structure for temporal reachability in the \emph{incremental} setting, where updates consist of inserting new edges or new labels on existing edges. Their structure has size $O(n^2\tau)$ and supports updates in amortized $O(n^2\log\tau)$ time, where $n$ is the number of vertices and $\tau$ is the \emph{lifetime} of the temporal graph, i.e., the number of distinct labels. Given vertices $u,v$ and times $t,t'$, it can determine whether there exists a temporal path from $u$ to $v$ departing no earlier than $t$ and arriving no later than $t'$. Queries require $O(\log\tau)$ worst-case time, and a witness path can also be reported efficiently.

Another related problem is to maintain the \emph{time-transitive closure}, namely the directed graph containing an arc $(u,v)$ whenever $v$ is temporally reachable from $u$. In~\cite{BarjonCCJN14}, this problem was studied in the restricted \emph{chronological incremental} setting, where inserted labels must be non-decreasing over time.

More recently, fully dynamic temporal reachability has been studied for restricted graph classes. In particular, Bil\`o, Gual\`a, Leucci, Proietti, and Straziota~\cite{BiloG0PS24} considered temporal forests and designed linear-space data structures supporting insertions and deletions of edge labels, together with temporal reachability, earliest arrival, and latest departure queries. Their main structure is also dynamic with respect to the topology of the forest: besides label updates, it supports insertion and deletion of singleton vertices, as well as link and cut operations that merge or split rooted trees. Updates are handled in $O(\log M)$ worst-case time, where $M$ is the total number of time labels associated with all the edges of the forest. Temporal reachability queries are answered in $O(\log M)$ worst-case time, while earliest arrival and latest departure queries require $O(\log L \cdot \log M)$ time, where $L$ is the maximum number of labels on a single edge. For temporal paths with fixed topology, the latter bound improves to $O(\log M)$. The same asymptotic guarantees were also extended to temporal forests with uniform latencies, while for arbitrary latencies incremental and decremental variants with $O(\log M)$-amortized update bounds were obtained. More precisely,  the incremental scenario supports the addition of new labels to existing edges, the addition of singleton vertices, and the link operation, while the decremental scenario only involves label/singleton deletions and cut operations.

Related trade-offs between space and query time were later investigated by Bil\`o, Gual\`a, Leucci, Proietti, and Straziota~\cite{BiloGLPS25} for \emph{fastest path oracles} in static temporal trees. They presented, for any parameter $t \geq 1$, a data structure of size $O(M + M^2/t^2)$ that answers fastest-path queries in $O(t \log(1+L))$ time, where $L$ is the maximum number of labels on an edge and $M$ is the total number of labels in the tree. Their queries can also enforce a prescribed time window. Moreover, they proved conditional lower bounds showing that these trade-offs are essentially optimal, up to polylogarithmic factors. The same work also provided evidence that extending such oracle results to general temporal graphs is significantly more difficult, even for temporal reachability queries without time constraints.









\subparagraph{Our results} In this paper, we design partially dynamic algorithms for maintaining earliest arrival times from a fixed source to all vertices in a temporal digraph with $n$ vertices, possibly with parallel arcs, where each arc has a time label and a positive latency (a traversal time of the arc).

For the incremental setting, in which updates consist of arc insertions, we give a data structure with constant query time and total update time $\Oh(n + m \log m)$, where $m$ is the total number of inserted arcs to an arc-less graph. For the decremental setting, where updates consist of arc deletions, we obtain the same query bound and total update time $\Oh(n + m \log m)$, where $m$ denotes the number of arcs in the initial graph. In both settings, once the earliest arrival time at a queried vertex $v$ is known, an earliest-arrival path from the source to $v$ can be reported in constant time per edge. We further consider the setting in which the departure-time threshold can be specified at query time. For this problem, we develop an incremental data structure supporting earliest-arrival queries in $O(\log m)$ time, with $O(n+\Delta_s m\log m)$ total update time and $O(n+m)$ space, where $\Delta_s$ denotes the number of distinct departure times of arcs leaving the source. The linear-space bound improves upon the straightforward approach of maintaining a separate data structure for each departure-time threshold.

We also prove conditional lower bounds under the Online Boolean Matrix-Vector Multiplication conjecture (a.k.a. the OMv conjecture). In both the incremental and decremental settings, there is no (possibly randomized) algorithm with amortized update time $\Oh(n^{1-\varepsilon})$ and amortized query time $\Oh(n^{2-\varepsilon})$, for any constant $\varepsilon>0$, even if we assume uniform latencies on the arcs. Therefore, up to these trade-offs, efficient total update time is essentially the strongest guarantee achievable. For the setting with arbitrary departure-time thresholds, we additionally prove an OMv-based conditional lower bound showing that our incremental algorithm is optimal up to logarithmic factors for dense temporal graphs. Our conditional lower bounds hold also for the case of undirected temporal graphs.

\subparagraph{Organization of the paper} The next section provides some preliminary notions. In \Cref{sec:incremental_simple} we present our algorithm for the incremental setting and in \Cref{sec:decremental} our algorithm for the decremental one. In \Cref{sec:lowerbound} we present our results about lower bounds for reachability problems in temporal graphs.
\section{Preliminaries}

A temporal digraph $G=(V,A)$ consists of a set $V$ of $n$ vertices and a set $A$ of $m$ temporal directed arcs, or simply arcs from now on. Each arc in $A$ is defined as a triple $(vu,t,l) \in A$, where $v$ and $u$ are two distinct vertices of $G$, while $t$ and $l$ are two numbers representing the exact time at which we can start traversing the arc from $v$ to $u$ and the traversal time of the arc (or latency), respectively. We assume that the latencies 
are positive, i.e., it takes a positive time to traverse an arc. Note that we do not assume that $G$ is a simple digraph; i.e., we allow many arcs with various time labels and latencies to appear in our temporal digraph. For the sake of simplicity, in the rest of the paper we assume that there cannot be two distinct arcs $(vu,t,l)$ and $(v'u',t',l')$ such that $vu=v'u'$, $t=t'$, and $l=l'$. All of our algorithms can be modified to also handle this case, by adding a unique identifier to each arc. 


A {\it temporal path} $P$  is a sequence of arcs with their time labels and latencies \[(v_0v_1, t_1, l_1),\ (v_1v_2, t_2, l_2), \dots, (v_{k-1}v_k, t_{k}, l_{k})\] such that for every $i=1, \dots, k-1$ we have $t_i + l_i \leq t_{i+1}$. We call $t_1$ the {\it departure time} of $P$ and $t_{k} + l_{k}$ the {\it arrival time} of $P$. The departure time $t_1$ of $P$ is the time at which we depart from $v_0$ in $P$, while the arrival time $t_k+l_k$ of $P$ is the time at which we arrive at $v_k$. So, in the rest of the paper, we will say also that $P$ is a path with departure time (resp., arrival time) of $t_1$ (resp., $t_k+l_k$) from $v_0$ (resp., $v_k$). 

Given two vertices $u$ and $v$ and a temporal interval $[t,t']$, with $t'\geq t$, we say that \emph{$v$ is reachable from $u$ within $[t,t']$} if there exists a temporal path from $u$ to $v$ whose departure time and arrival time are both contained in the interval $[t,t']$. 
When there exists $t \in \mathbb{N}$ such that $v$ is reachable from $u$ within $[0,t]$, we simply say that $v$ is reachable from $u$. 
Moreover, we say that $t'$ is \emph{the earliest arrival time from $u$ to $v$ w.r.t. $t$} if $v$ is reachable from $u$ in the interval $[t,t']$, but $v$ is not reachable from $u$ in the time interval $[t,t')$. When $t=0$, we will simply say that $t'$ is the earliest arrival time from $u$ to $v$.


\subparagraph{Balanced binary search trees.}
All our algorithms use balanced binary search trees storing arcs in the form of tuples $a=(vu,t,l)$, keyed by some function of $t$ and $l$, denoted as $\key(a)$.  
A balanced binary search tree $S$, implemented for instance as an AVL tree, supports the following operations: 
\begin{itemize}
    \item $S.\mathtt{Initialize}()$: initializes the structure with an empty set of arcs;
    \item $S.\mathtt{Empty}()$: returns $\mathtt{true}$ if there is no arc in $S$ and $\mathtt{false}$ otherwise.
    \item $S.\mathtt{Insert}(a)$: inserts arc $a$ into $S$;
    \item $S.\mathtt{Remove}(a)$: removes arc $a$ from  $S$;
    \item $S.\mathtt{Successor}(k)$: finds the arc $a\in S$ with the smallest key $\key(a)$ such that $\key(a) \geq k$; if there is no such arc, return $\perp$;
    \item $S.\mathtt{Predecessor}(k)$: finds the arc $a \in S$ with the greatest key $\key(a)$ such that $\key(a) < k$; if there is no such arc, return $\perp$;
    \item $S.\mathtt{Smallest}()$: returns the arc with the smallest key in $S$; if $S$ is empty, return $\perp$;
    \item $S.\mathtt{Next}(a)$: returns an arc $a' \in S$ that is a direct successor of $a$ in the order of keys; if there is no such successor, return $\perp$.
\end{itemize}
Implementation as AVL trees supports the constant time for $\mathtt{Initialize}()$ operation and $\Oh(\log |S|)$ time for other operations, while also providing linear space.

\section{Incremental algorithm}\label{sec:incremental_simple}

In this section, we design an incremental data structure that is capable of answering single-source earliest-arrival-time queries in constant time. The data structure is also capable of reporting an earliest arrival temporal path from the source to the queried vertex in constant time per arc. 
We observe that the same construction allows us to report earliest-arrival times from the source to $v$ w.r.t. some fixed value $t$, for every vertex $v$ as we can simply ignore the addition of the arcs whose corresponding time labels are strictly smaller than $t$ as these arcs cannot be used in any temporal path whose departure time is $t$ or larger. 

\begin{theorem}\label{thm:inc}
    Let $G$ be a temporal arc-less digraph with $n$ vertices and let $s$ be a  vertex of $G$. There is an incremental data structure that maintains earliest arrival paths from $s$ to every other vertex of $G$ by supporting the following operations:
    \begin{itemize}
        \item $\Init()$ -- initializes the data structure with a temporal digraph $G$ having no arcs in $\mathcal{O}(n)$ time;
        \item $\AddArc(vu,t,l)$ -- adds arc $(vu,t,l)$ to $G$; the total time of adding $m$ arcs is $\Oh(n + m\log m)$;
        \item $\Dist(v)$ -- returns the earliest arrival time from $s$ to $v$ in the current digraph $G$, or $\infty$ if $v$ is not reachable from $s$, in $\Oh(1)$ worst case time;
        \item $\Path(v)$ -- returns an earliest arrival path from $s$ to $v$ in the current digraph $G$, or $\bot$ if $v$ is not reachable from $s$, in time linear in the size of the returned path.
    \end{itemize}
\end{theorem}


It is well known that there exists a directed tree $T$ rooted at $s$, a.k.a. \emph{earliest-arrival tree}, such that, for every vertex $v$ that is reachable from $s$ in $G$, the unique path from $s$ to $v$ in $T$ is an earliest arrival path. The idea is to dynamically maintain an earliest-arrival tree rooted at $s$.

The data structure internally maintains the following invariant information for every vertex $v$ of $G$:
    \begin{itemize}
        \item $\ea[v]$ -- the 
        earliest arrival time from $s$ to $v$ in the current temporal digraph $G$ or $\infty$ if $v$ is not reachable from $s$ in $G$. 
        \item $\pred[v]$ -- the arc $(uv,t,l)$ preceding $v$ on its earliest arrival path in the earliest-arrival tree in the current temporal digraph $G$ rooted at $s$, or $\bot$ if $v$ is not reachable from $s$ in $G$ or $v=s$;
        \item $\mathtt{N}[v]$ -- the set of all arcs outgoing from $v$ whose time labels are strictly smaller than $\ea[v]$ w.r.t. the current temporal digraph $G$.  
    \end{itemize}
    All the $\ea$ and the $\pred$  values are stored in an array, while each $\mathtt{N}[v]$ is stored as an AVL tree whose keys are the time labels. 
    
    \subparagraph{Initialization.} We assume that $G$ initially has no arcs. As no vertex apart from $s$ is reachable from $s$, we set $\ea[v]=\infty$ for every $v\in V\setminus \{s\}$ and $\ea[s]=0$. Moreover, $\pred[v]=\bot$, for every vertex $v \in V$. As there are no arcs, all $\mathtt{N}[v]$'s are initialized to be empty AVLs. This requires time $\Oh(1)$ per vertex. Therefore, the initialization of the data structure requires $\Oh(n)$ time. We observe that all the invariants are maintained.
    
\begin{figure}[]
        \begin{center}
        \includegraphics[scale=0.82]{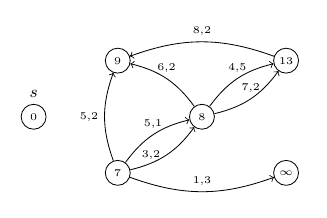}
        \includegraphics[scale=0.82]{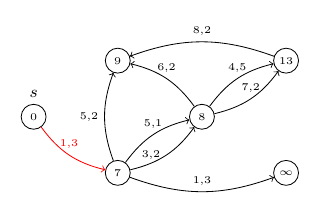}
        \includegraphics[scale=0.82]{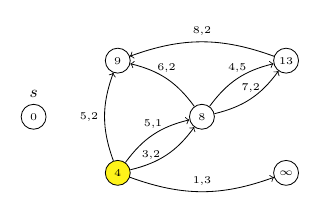}

        \vspace{-0.5cm}
        \includegraphics[scale=0.82]{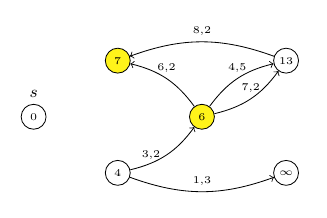}
        \includegraphics[scale=0.82]{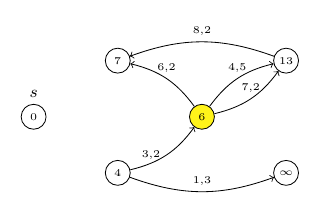}
        \includegraphics[scale=0.82]{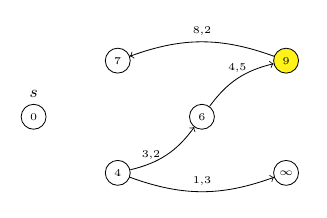}
        
        \vspace{-0.2cm}
        \includegraphics[scale=0.82]{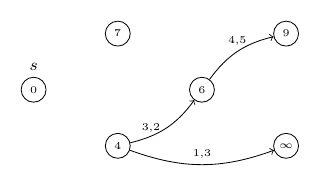}
        \caption{The evolution of the data structure after adding the red arc. The number shown inside each vertex $v$ is $\ea[v]$ while the arcs outgoing from $v$ are those stored inside  $\mathtt{N}[v]$. The numbers associated with each arc are respectively its time label and latency. For clarity, we omit here the information about the $\pred$. When the red arc is added, we call our procedure to add arcs. Vertices colored yellow are the ones belonging to the queue $Q$ at the moment.}
        \label{fig:incremental_ilustration}
        \end{center}
        \end{figure}

    \subparagraph{Adding an arc.} The {$\AddArc(vu, t, l)$} procedure is presented in \Cref{alg:incremental} and in \Cref{fig:incremental_ilustration}; here we describe it in more detail and argue about its correctness and its running time.

    
    We check whether $t < \ea[v]$. If this is the case, the added arc $(vu,t,l)$ cannot be used at the moment by a temporal path, as the earliest arrival time at $v$ is currently higher than the time label of this arc. However, as the values $\ea[v]$'s are monotonically non-increasing when we add new arcs to the graph, we need to keep track of the arc $(vu,t,l)$, so we add it to $\mathtt{N}[v]$ and return. This arc cannot be included now in any temporal path starting from $s$. 
    
    If $\ea[v] \leq  t$, we first check whether the value $\ea[u]$ can be improved: if $\ea[u]> t+ l$ then we set $\ea[u]=t+l$ and create a queue $Q$ that contains $u$; otherwise the procedure ends. 
    
    Then as long as there is a vertex in $Q$, we pick any of them from $Q$, say it $w$, and immediately remove $w$ from $Q$. Now we iterate through all arcs belonging to $N[w]$ with time labels greater than or equal to $\ea[w]$. More precisely we  {\it check} them in the following way: for every arc $(wz,t',l') \in N[w]$, if $\ea[z]> t' + l'$, we set $\ea[z]=t'+l'$, $\pred[z]=(wz,t',l')$ and we add $z$ to $Q$. The arc $(wz,t',l')$ is deleted from $N[w]$.

    Our procedure ends when $Q$ becomes empty. Note that improving the value of $\ea$ of one vertex may cause further changes in the structure, and more vertices may be affected as a consequence.
    
    

    We now show that, at the end of the procedure, all the invariants of our data structure are maintained. 
    Note that the values $\ea[v]$ are monotonically non-increasing, and that when the value $\ea[v]$ decreases, $v$ is always added to $Q$ and $N[v]$ is modified accordingly. So, for every vertex $v$ not in $Q$, the set $N[v]$ contains exactly the arcs of the current temporal digraph $G$ whose corresponding time labels are strictly smaller than $\ea[v]$.

    We now show that, for every vertex $v$, $\ea[v]$ is the earliest arrival time at $v$ from $s$ in the current temporal digraph. 
    For the sake of contradiction, suppose that there exists a vertex $v$ such that $v$ is reachable by time $t$ from $s$ in the current temporal digraph $G$ and $t<\ea[v]$. Let $T$ be an earliest-arrival tree rooted at $s$ for the current temporal digraph $G$. Among all possible choices of $v$, we pick the vertex $v$ that is closest to $s$ in $T$ w.r.t. its hop distance from $s$. Let $x$ be the predecessor of $v$ in $T$ and let $(xv,t',l')$ be the arc in $T$. By the choice of $v$, $\ea[x]$ equals the earliest arrival time at $x$. Moreover, $\ea[v] > t' + l'$. As $t' \geq \ea[x]$, the arc $(xv,t',l')$ cannot belong to $N[x]$. Thus this arc must have already been checked in our procedure and thus $\ea[v] \leq t' + l'$, which is a contradiction. Note that $\pred[x]$ for any vertex $x$ is also updated accordingly. Thus, all invariants are maintained after the completion of the $\AddArc$ procedure.

    \begin{algorithm}[]
    \SetKwInOut{Input}{Input}
    \SetKwInOut{Output}{Output}
    \underline{function} $\AddArc(vu, t, l)$\;
    \Input{An arc $(vu,t,l)$}
    \If{$t<\ea[v]$} 
    {
        $\mathtt{N}[v].\mathtt{Insert}(vu, t, l)$\;
        \Return;
    }
    $Q.\mathtt{Init}()$\;
    \If{$\ea[u] > t + l$}
    {
        $\ea[u] \leftarrow t+l$\;
        $\pred[u] \gets (vu,t,l)$\;
        $Q.\mathtt{push}(u)$\; 
    }
    \While{$Q \neq \emptyset$}
      {
        $w \gets Q.\mathtt{Pop}()$\;
        \While{$\mathtt{N}[w].\mathtt{Successor}(\ea[w]) \neq \bot$}
        {
            $(wz, t', l') \leftarrow \mathtt{N}[w].\mathtt{Successor}(\ea[w])$\;
            $\mathtt{N}[w].\mathtt{Remove}(wz, t', l')$\;
            \If{$t' + l' < \ea[z]$}
            {
                $\ea[z] \leftarrow t' + l'$\;
                $\pred[z] \gets (wz,t',l')$\;
                $Q.\mathtt{Push}(z)$\;
            }
        }
      }
    \caption{The procedure for adding arcs. The set $Q$ is here initialized to be an empty queue.}
    \label{alg:incremental}
    \end{algorithm}
        
    \subparagraph{Queries.} When we query for the earliest arrival time at $v$ in the current temporal digraph $G$, the query algorithm returns $\ea[v]$. This operation requires time $\Oh(1)$. Moreover, if $\ea[v]\neq \infty$, the earliest arrival path from $s$ to $v$ stored in the data structure can be returned in asymptotical linear time w.r.t. its number of edges using the array $\pred$.
    
    \subparagraph{Analysis of the total update time.} Let $m$ be the overall number of $\AddArc$ operations throughout the lifetime of the data structure. Observe also that an arc $(vu,t,l)$ is added at most once into the AVL tree $\mathtt{N}[v]$ and never added to the AVL trees of other vertices. Therefore, the total number of edges stored in all the AVL trees is $\Oh(m)$. Hence, each operation on the AVL $\mathtt{N}[v]$ takes $\Oh(\log |\mathtt{N}[v]|) = \Oh(\log m)$ time. Each of the arcs stored in the AVL trees is checked at most once throughout the lifetime of the procedure, as it is only checked after being deleted from the AVL tree containing it. This means that each arc is checked at most once, for a total number of $\Oh(m)$ checks. Thus we need total time $\mathcal{O}({m\log m})$ for all operations $\AddArc$.

\section{Decremental algorithm}\label{sec:decremental}

In this section, we design a decremental data structure that is capable of answering single-source earliest-arrival-time queries in constant time. The data structure is also capable of reporting an earliest arrival temporal path from the source to the queried vertex in constant time per arc. As already observed in~\Cref{sec:incremental_simple}, the same construction allows us to report single-source earliest arrival times w.r.t. some fixed value $t$. Indeed, we can preprocess the digraph by removing all arcs having a time label strictly smaller than $t$ from it. 
More precisely, we prove the following theorem.


\begin{theorem}
Let $G$ be a temporal digraph with $n$ vertices and $m$ arcs, and let $s$ be a vertex of $G$. There is a decremental data structure that maintains earliest arrival times from $s$ to every other vertex of $G$ by supporting the following operations:
\begin{itemize}
    \item $\Init()$ -- initializes the data structure with the input temporal digraph $G$ in $\Oh(n+m\log m)$ time;
    \item $\RemoveArc(vu, t, l)$ -- removes the arc $(vu,t,l)$ from $G$; the total time of removing all $m$ arcs is $\Oh(n + m\log m)$;
    \item $\Dist(v)$ -- returns the earliest arrival time from $s$ to $v$ in the current digraph $G$, or $\infty$ if $v$ is not reachable from $s$, in $\Oh(1)$ worst case time;
    \item $\Path(v)$ -- returns an earliest arrival path from $s$ to $v$ in the current digraph $G$, or $\bot$ if $v$ is not reachable from $s$, in time linear in the size of the returned path.
\end{itemize}
\end{theorem}
The data structure maintains the following information satisfying the invariants:
\begin{itemize}
    \item $\ea[v]$ -- the earliest arrival time from $s$ to $v$ in the current temporal digraph $G$ or $\infty$ if $v$ is not reachable from $s$ in $G$. 
    \item $\mathtt{parent}[v]$ -- the arc $(uv,t,l)$ preceding $v$ on its earliest arrival path in the earliest arrival tree in the current temporal digraph $G$ rooted at $s$, or $\bot$ if $v$ is not reachable from $s$ in $G$ or $v=s$;
    \item $\mathtt{N}_{in}[v]$ - the set of all arcs $(uv,t,l)$ entering $v$ for which $t+l\geq \ea[v]$ such that there exists a temporal path from $s$ to $v$ using $(uv,t,l)$; 
    \item $\mathtt{N}_{out}[v]$ - the set of all arcs leaving $v$ whose time labels are greater than or equal to $\ea[v]$.
\end{itemize}

 All the $\ea$ and the $\mathtt{parent}$ values  are stored in an array and there are two distinct AVL trees per vertex $v$ of $G$, one for $\mathtt{N}_{in}[v]$ and one for $\mathtt{N}_{out}[v]$.
The keys of $\mathtt{N}_{out}$ are the time labels triples , while the keys of $\mathtt{N}_{in}[v]$ are the sums of time label and latency.

\subparagraph{Initialization.} The data structure starts with our initial temporal digraph $G=(V,A)$ that is given as input. First, we use \Cref{alg:incremental} we designed for the incremental scenario to compute  the values of $\ea[v]$ and $\pred[v]$ for every vertex $v \in V$. Then we scan all the arcs $(uv,t,l)$ of $G$ to initialize all AVL trees correctly. More precisely, we add the arc to $\mathtt{N}_{out}[u]$ and to $\mathtt{N}_{in}[v]$ if $t\geq \ea[u]$ (then we also have $t+l\geq \ea[v]$).
Note that each arc is stored in at most two AVL trees.  
There can be arcs which are stored in no such sets. Then we know that there are no temporal paths from $s$ using them, as they have too small time labels.
Moreover, through the lifetime of the structure, the value of $\ea[v]$ for any vertex $v$ is monotonically non-decreasing; thus there will be no temporal paths using the edges we already discarded during initialization. As we checked each arc, all invariants above are maintained. The total time required for the initialization is $\Oh(n + m\log m)$, as this is the time required to set the value of all $\ea$s and later to add arcs to appropriate AVL trees.

        \begin{figure}[t]
        \begin{center}
            
        \includegraphics[scale=0.82]{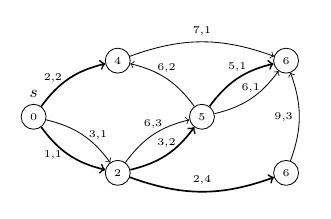}
        \includegraphics[scale=0.82]{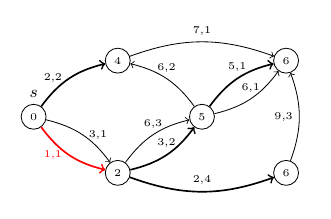}
        \includegraphics[scale=0.82]{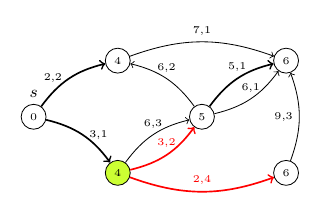}

        \vspace{-0.5cm}
        \includegraphics[scale=0.82]{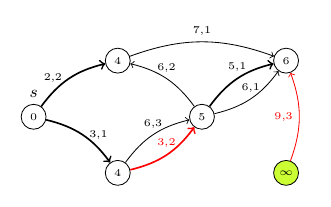}
        \includegraphics[scale=0.82]{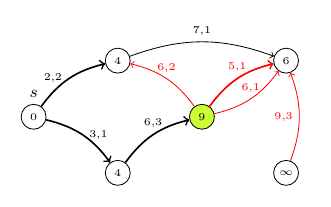}
        \includegraphics[scale=0.82]{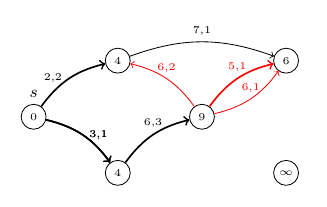}

        \vspace{-0.5cm}
        \includegraphics[scale=0.82]{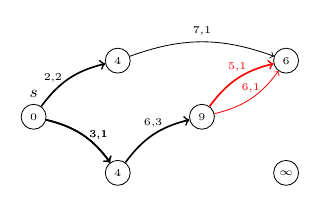}
        \includegraphics[scale=0.82]{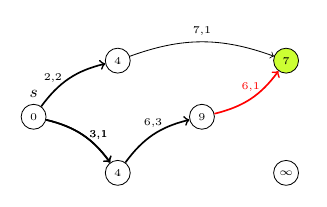}
        \includegraphics[scale=0.82]{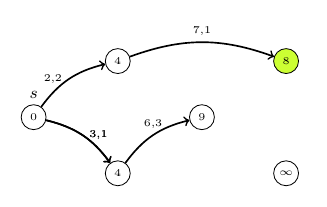}

        \vspace{-0.5cm}
        \includegraphics[scale=0.82]{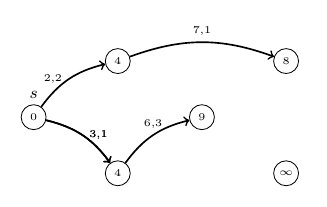}
        \caption{The evolution of the data structure after removing the red arc. The number shown inside each vertex $v$ is $\ea[v]$ and the arcs ingoing and out going belong to $\mathtt{N}_{in}[v]$ and $\mathtt{N}_{out}[v]$ respectively. The incoming thick arc is a $\pred[v]$. Red edges are marked to be removed and green vertices have just their value of $\ea$ changed.}
        \label{fig:decremental_ilustration}
        \end{center}
        \end{figure}
    
\subparagraph{Queries.} For every vertex $v$, our data structure maintains the invariant that $\ea[v]$ is the earliest arrival time such that vertex $v$ is reachable from $s$ in the current temporal digraph $G$. Thus, given vertex $v$, we return $\ea[v]$. This operation requires $\Oh(1)$ time. Moreover, if $\ea[v]\neq \infty$, the earliest arrival path from $s$ to $v$ stored in the data structure can be returned in asymptotical linear time w.r.t. its number of arcs using the array $\mathtt{parent}$.

\begin{algorithm}[ht]
\SetKwInOut{Input}{Input}
\SetKwInOut{Output}{Output}
\underline{function} $\RemoveArc(uv, t, l)$\;
\Input{An arc $(uv,t,l)$}
    $\mathtt{N}_{out}[u]$.\algtt{Remove}$(uv, t, l)$\;
    $\mathtt{N}_{in}[v]$.\algtt{Remove}$(uv, t, l)$\;
    \If{$(uv,t,l) \neq \pred[v]$} 
    {
        \Return;
    }
    \If{$\mathtt{N}_{in}[v].\algtt{Empty}() = \mathtt{true}$ } 
    {
        $\pred[v] \leftarrow \bot$\;
        $\ea[v] \leftarrow \infty$\;
        \While{$\mathtt{N}_{out}[v].\mathtt{Empty}() = \mathtt{false}$}
        {
            $\RemoveArc(\mathtt{N}_{out}[v].\algtt{Smallest}())$\;
        }
        \Return;
    }
    $(wv, t', l') \leftarrow \mathtt{N}_{in}[v].\mathtt{Smallest}()$\;
    $\pred[v]\leftarrow (wv,t',l')$\;
    $\ea[v] = t' + l'$\;

    \While{$\mathtt{N}_{out}[v].\mathtt{Predecessor}(\ea[v]) \neq \bot$}
    {
        $\RemoveArc(\mathtt{N}_{out}[v].\algtt{Smallest}())$\;
    }    
    \caption{The procedure for removing arcs.}
    \label{alg:decremental}
    \end{algorithm}

\subparagraph{Removing an arc.} The $\RemoveArc(uv, t, l)$ procedure is presented in \Cref{alg:decremental} and in \Cref{fig:decremental_ilustration}; here we describe it in more detail. First, we remove the arc $(uv, t, l)$ from $\mathtt{N}_{out}[u]$ and $\mathtt{N}_{in}[v]$. Next, we check $u$ is connected to $v$ by $\pred[v]$. If this is not the case, we finish the procedure here. 

If $\mathtt{N}_{in}[v]$ is empty now, the vertex $v$ is no longer reachable from the source $s$. Thus we set $\pred[v]=\bot,\ \ea[v]=\infty$ and we call $\RemoveArc$ operation for all arcs belonging to $\mathtt{N}_{out}[v]$.

If $\mathtt{N}_{in}[v]$ is not empty, we pick the arc $wv$ with the smallest sum of the time label $t$ and the latency $l$ from the set $\mathtt{N}_{in}[v]$. Then we update $\pred[v]=(wv,t,l)$, set $\ea[v]=t+l$. We need now to update all vertices $x$ for which the time label of the arc $vx$ is smaller than current $\ea[v]$. Therefore we perform $\RemoveArc$ operation on all arcs belonging to $\mathtt{N}_{out}[v]$ with time labels smaller than $\ea[v]$. 

Note that similarly to the incremental case, removing one arc may lead to the removal of many edges as the consequence. 

First, note that after removal of each edge and each change to $\ea[v]$ for any vertex $v$, we always update $\mathtt{N}_{in}[v]$ and $\mathtt{N}_{out}[v]$. With this change we also ensure that the equality $\ea[v] = t + l$ for some arc $a$ with time label $t$ and latency $l$ with $a$ being $\pred[v]$.

We need to show that for every vertex $\ea[v]$ is indeed the earliest arrival time at which $v$ is reachable from $s$ starting at time $t_0$. Suppose not, and let $w$ be a vertex with the smallest earliest arrival time among such vertices; we know that its earlier arrival time is strictly smaller than $\ea[w]$. Let $P$ be a path witnessing the arrival time in $w$ smaller than $\ea[w]$. We can assume that $P$ witnesses the earliest arrival time for every vertex on it, but $w$. Let $x$ be a predecessor of $w$ on $P$ and let $xw$ be the arc on $P$ with time label $t'$ and latency $l'$. By the assumption, we know that $\ea[x]$ is the earliest arrival time that we can reach $x$ from $s$ starting at time $t_0$ and that $\ea[x]<\ea[w]$. Thus the earliest arrival time that we can reach $w$ from $s$ starting at time $t_0$ is equal to $t'+l' < \ea[w]$. 
As we know the invariant was fulfilled after the initialization step, we know that $\ea[w]$ must have changed. Let $\ea'[w]$ be the value of $\ea[w]$ just after the initialization step. We have $\ea'(w)\leq t'+l'$, as by the assumption $\ea'[w]$ is the earliest arrival time that we can reach $w$ in the starting graph. Thus the arc $xw$ belonged to $\mathtt{N}_{in}[w]$. As now we have $t'+l'< \ea[w]$, some arc adjacent to $w$ was deleted. In that step we updated $\ea[w]$ for the smallest value inside $\mathtt{N}_{in}(w)$, which is $\leq t'+l'$. Thus we get a~contradiction.

\subparagraph{Analysis of the total update time.} We recall that $m$ is the number of arcs in the initial temporal digraph $G$. Then all operations on sets $\mathtt{N}_{in}[v]$ and $\mathtt{N}_{out}[v]$ at any moment take at most $\Oh(\log m)$ time. Each arc is considered and removed at most once in the lifetime of the structure, thus we need in total $\Oh(m\log m)$ operations.


\section{Bounded time window}

In this section we present a refined incremental algorithm that reports earliest arrival times from a source to every other vertex, where we can additionally provide the departure time $t_0$ from $s$ to the query, instead of having it fixed for the lifetime of the data structure. The refined algorithm incurs an extra factor of $\Delta_s$ (the final number of possible departure times from the source) in the total update running time. In Section~\ref{sec:lowerbound} we show that this extra factor is likely unavoidable. The precise theorem that we prove is the following.


\begin{theorem}\label{thm:boundedTimeWindow}
    Let $G=(V,A)$ be a temporal digraph on $n$ vertices, let $s$ be a source vertex. There is an incremental data structure maintaining $G$ that offers the following arc updates and queries: 
    \begin{itemize}
        \item $\Init()$ -- initializes the data structure with an empty temporal digraph $G$ on $n$ vertices in $\mathcal{O}(n)$ time;
        \item $\AddArc(vu,t,l)$ -- adds arc $(vu,t,l)$ to $G$; the total time of adding $m$ arcs is $\mathcal{O}(n + \Delta_s m \log m)$, where $\Delta_s$ stands for the number of different time labels among the outgoing arcs of $s$ in the final graph;
        \item $\Dist(v,t_0)$ -- returns the earliest arrival time from $s$ to $v$ with respect to $t_0$. The worst case time is $O(\log m)$.
        \item $\Path(v,t_0)$ -- returns an earliest arrival path from $s$ to $v$ with respect to $t_0$, or $\bot$ if $v$ is not reachable from $s$, in time $O(\log m)$ plus linear time in the size of the returned path.
    \end{itemize}
    The data structure uses $\mathcal{O}(n+m)$ space.
\end{theorem}

\begin{proof}
Let us denote, for every vertex $v$,
by $A^+_v$ the set of arcs currently leaving $v$, and by $A^-_v$ the set of arcs currently entering $v$. 
For every vertex $v$, we also denote by $D_v=\{t \mid (vu,t,l) \in A^+_v\}$ the set (without repetitions) of possible departure times from $v$ and by $R_v=\{t+l \mid (uv,t,l) \in A^{-}_v\}$ the set of possible arrival times at $v$. All of the above sets change as the updates arrive, so they always refer to the current state of the temporal graph $G$. For an arc $a=(uv,t,l)$, let $\arr(a)$ denote $t+l$. 
Over the sequence of updates, the algorithm maintains the following information:
\begin{itemize}
        \item For each arc $a \in A$ we store  $\ld[a] \in D_s \cup \{ -\infty \}$ -- the 
        latest departure time $t$ from $s$ such that there is a temporal path whose last arc is $a$ and whose departure time from $s$ is $t$; 
        \item For every arc $a \in A$ we store $\pred[a]$ -- the predecessor of $a$ on the path 
        starting from $s$ at time $\ld[a]$, or $\bot$ if 
        $\ld[a]=-\infty$ or $a$ is the first edge of such path. 
        \item For each vertex $v$ we store
        $\inedges[v]$ -- the balanced binary search tree storing $A^{-}_v$, keyed by $\arr(a)$; every node is augmented with maximum $\ld$ in the subtree; this allows querying for the arc with maximum $\ld$ of all the arcs $a \in A^{-}_v$ with $\arr(a) \leq t$, let us denote this query by $\prefld(t)$. Whenever the value $\ld[a]$ of an arc $a$ changes, the augmented information in the corresponding tree is updated in $O(\log m)$ time. 
        \item For every vertex $v$ we store $\outedges[v]$ -- the balanced search tree storing $A^{+}_v$, keyed by time labels.
        
    \end{itemize}

\begin{algorithm}[t]
\caption{Insertion of an arc and propagation.}\label{alg:addBW}
\SetKwInOut{Input}{Input}
\SetKwInOut{Output}{Output}
\underline{function} $\AddArc(vu, t, l)$\;
\Input{An arc $a=(vu,t,l)$}
  $\outedges[v].\mathtt{Insert}(a)$ \tcp*{ keyed by $t$} 
  $\inedges[u].\mathtt{Insert}(a)$ \tcp*{ keyed by $t+l$} 
  \eIf{$v=s$}{$\ld[a]\gets t$; 
  $\pred[a]\gets\bot$\;}
  {$b = \inedges[v].\prefld(t)$
  \tcp*{$b \in \inedges[v]$: $\arr(b)\le t$ and $\ld[b]$ is maximum} 
   \eIf{$b \neq \bot$}{
   $\ld[a]\gets\ld[b]$}{$\ld[a] \gets -\infty$}  
   $\pred[a]\gets b$\;}
  $Q \gets\{a\}$\;
  \While{not $Q.\mathtt{Empty()}$}{
    $b=(xy,t_b,l_b) \gets Q.\mathtt{Pop()}$; $d\gets\ld[b]$ \tcp*{current value}
    \lIf{$y=s$ \textbf{or} $d=-\infty$}{\textbf{continue}}
    $a'\gets \outedges[y].\mathtt{Successor}(\arr(b))$\;
    \label{ln:succ}
    
    \While{$a'\neq\bot$ \textbf{and} $\ld[a']<d$}{\label{ln:loop}
      $\ld[a']\gets d$; $\pred[a']\gets b$\;\label{ln:set}
      \lIf{$a'\notin Q$}{$Q.\mathtt{Push(a')}$}
      $a'\gets \outedges[y].\mathtt{Next}(a')$\;
    }
  }
\end{algorithm}
\subparagraph{Initialization.} We initialize on an edgeless graph $G$, so no values $\ld$ are assigned to the arcs upfront. Balanced trees $\inedges[v]$ and $\outedges[v]$ are initialized as empty for every vertex $v \in V$. Thus, the total initialization time is $\mathcal{O}(n)$.
\subparagraph{Query.} Upon query $\Dist(v,t_0)$, the data structure inspects $\inedges[v]$ sorted by arrival times. The goal is to find the earliest arc $a$ in $\inedges[v]$ whose $\ld[a] \geq t_0$. This can be done by traversing the tree $\inedges[v]$ from the root, and following the leftmost path towards the leaf such that $\ld$ in the subtree is at least $t_0$. Such a search takes worst case time of $O(\log |A^{-}_v|)$. For the $\Path(v,t_0)$ query, once $a$ is located in $O(\log |A^{-}_v|)$ time, we follow 
the parent pointers to reconstruct the path.
\subparagraph{Update.} The $\AddArc(vu,t,l)$ operation is presented in Algorithm~\ref{alg:addBW}, and we describe it here in more detail. When a new arc $a=(vu,t,l)$ is added to $G$, the algorithm first updates $\outedges[v]$ and $\inedges[u]$ accordingly by inserting $a$. Clearly,
if $v=s$ then we can set $\ld[a]=t$, as to use arc $a$ in a one edge temporal path we have to depart from $s$ at time $t$. We set $\pred[a]=\bot$ as this is the first edge of all temporal paths using this edge.

Otherwise, if $v \neq s$, we have to find the value $\ld[a]$ and assign it. For that, we consider all edges $b \in A^{-}_v$ with $\arr(b) \leq t$, so that they can be combined in a temporal path with arc $a$. Among those, we choose the one with maximum $\ld$ value by issuing $\prefld(t)$ query. We assign the $\ld$ value of this edge to $\ld[a]$. In the case $\prefld(t)$ returns $\bot$, the arc $a$ cannot be reached from the source so we set $\ld[a]=-\infty$.
We update $\pred[a]$ accordingly.

Next, we need to change $\ld$ values of other arcs that may have been affected by the insertion of the new arc $a$. We add $a$ to the FIFO queue $Q$, which stores the arcs whose $\ld$ value changed, and this change needs to be propagated further. We then process the arcs in $Q$ until $Q$ is empty, i.e., there is no further propagation needed. In each iteration we pop one arc $b=(xy,t_b,l_b)$ from $Q$. If the arc we popped is entering $s$, or $\ld[b]=-\infty$, then we do not need to propagate as $b$ does not affect any further arcs. Otherwise we inspect $\outedges[y]$ (sorted by time label). Observe that the edges that may directly profit from the change in $\ld[b]$ are the edges of $A^{+}_y$ whose time label is at least $\arr(b)$, as these are the edges that can follow $b$ in the temporal path. Thus, we let $a'$ be the first such edge, which we can find by issuing $\algtt{Successor}(\arr(b))$ query to $\outedges[y]$. Using $a'$ we iterate through $\outedges[y]$ as long as $\ld[a'] < \ld[b]$: in that case we improve $\ld[a']$ whose new value is $\ld[b]$. $\ld$ values are monotone with time labels, so once this stops being the case $b$ cannot affect any more edges of $\outedges[y]$ and we can stop the iteration. This completes the description of the procedure.

\subparagraph{Time analysis.} 
For the initialization we pay $\mathcal{O}(n)$ time.
For the analysis of the total update time consider that
every arc is pushed onto $Q$ at most $\Delta_s$ times. Each time an arc is pushed to $Q$, a constant number of balanced search tree operations are issued, each costing at most $\mathcal{O}(\log m)$ time. As a consequence, the total update time is $\mathcal{O}(n+\Delta_s m \log m)$. It was already argued that the worst case query time is $\mathcal{O}(\log m)$.  
\end{proof}

It is worth noting that the total update time of Theorem~\ref{thm:boundedTimeWindow} can also be achieved by maintaining $\Delta_s$ copies of the data structure of Theorem~\ref{thm:inc}, albeit at the cost of increased space usage.

Finally, in the chronological setting, where edges are inserted in increasing order of their time labels, the data structure of Theorem~\ref{thm:boundedTimeWindow} requires no propagation. Consequently, its total update time is $\mathcal{O}(n + m \log m)$, matching that of Theorem~\ref{thm:inc}.
\section{Lower bounds}\label{sec:lowerbound}

In this section we prove conditional lower bounds for the dynamic single-source temporal path reachability problem, i.e., the problem in which we only want to understand whether a queried vertex is reachable from the source by some temporal path (not necessarily an earliest-arrival temporal path) or not. In the following, we denote by $[n]$  the set $\{1,2,\dots, n\}$. Our reductions are based on the
\emph{Online Boolean Matrix-Vector Multiplication} (OMv) conjecture of
Henzinger, Krinninger, Nanongkai and Saranurak~\cite{OMv}. In OMv, one is
given an $n\times n$ Boolean matrix $M$ that may be preprocessed in polynomial
time. Afterwards, the algorithm receives Boolean column vectors
$v_1, v_2, \dots, v_n$ one by one and, before seeing $v_{i+1}$, must output
$Mv_i \in \{0,1\}^n$ for each $i\in[n]$. A straightforward algorithm runs in
$\Oh(n^3)$ total time after reading $M$. Although the problem admits an
algorithm with running time
$n^3/2^{\Omega(\sqrt{\log{n}})}$~\cite{larsen2017faster,williams2007matrix},
no $\Oh(n^{3-\varepsilon})$-time algorithm is known for any $\varepsilon > 0$. The
OMv conjecture asserts that no such algorithm exists.

\begin{conjecture}[OMv-conjecture, cf., Conjecture 1.1 in~\cite{OMv}]\label[conjecture]{con:omv}
    For any $\varepsilon > 0$, there is no algorithm with polynomial preprocessing
    time and total running time $\Oh(n^{3-\varepsilon})$ that solves OMv with
    error probability at most $1/3$.
\end{conjecture}

For our reductions, it is convenient to work with the \emph{OuMv problem}. In
this variant, we are additionally given Boolean row vectors
$u_1, u_2, \dots, u_n$, and the task is to compute
$u_iMv_i \in \{0,1\}$ for each $i\in[n]$ before seeing the next pair
$(u_{i+1}, v_{i+1})$.

\begin{lemma}[cf., Corollary 2.8 and Theorem 2.13 in~\cite{OMv}]\label[lemma]{lem:omv_to_oumv}
    Assuming the OMv conjecture, for any $\varepsilon > 0$ there is no algorithm
    with polynomial preprocessing time and
    $\Oh(n^{3-\varepsilon})$ computation time that solves OuMv with error probability
    at most $1/3$. This holds even if the initial
    matrix $M$ is symmetric.
\end{lemma}

All lower bounds below are stated for undirected temporal graphs and can be easily extended to work on temporal digraphs as well.

\begin{lemma}\label[lemma]{lem:partially_undirected_lowerbound}
    Assuming the OMv conjecture, for any $\eps > 0$ there is no partially dynamic
    algorithm solving the Single Source Temporal Path Reachability (SSTP) problem on
    undirected graphs with $n$ vertices, with:
    \begin{itemize}
        \setlength{\itemsep}{0pt}
        \item polynomial time preprocessing, 
        \item update in $\Oh(n^{1-\eps})$ amortized time, and
        \item query in $\Oh(n^{2-\eps})$ amortized time.
    \end{itemize}
    This lower bound holds even if the algorithm is allowed to output incorrect
    answers with probability at most $1/3$.
\end{lemma}
\begin{proof}
    We reduce OuMv to SSTP. The reduction uses $\Oh(n)$ vertices,
    $\Oh(n^2)$ updates, and $\Oh(n)$ queries. Hence, together
    with~\cref{lem:omv_to_oumv}, any SSTP algorithm with sublinear amortized
    update time and subquadratic amortized query time would yield an
    $\Oh(n^{3-\eps})$-time algorithm for OuMv. We first describe the
    incremental reduction and then its decremental analogue.

    \subparagraph{Incremental model.} We initialize the structure with a graph $G$ on
    $3n+1$ vertices: $s$, $x_1,x_2, \dots, x_n$, $y_1, y_2, \dots, y_n$,
    $t_1, t_2, \dots, t_n$. For every $i,j\in [n]$ such that $M[i][j]=1$, we
    add temporal edges $(x_i, y_j)$ with time labels $1,2,\dots, n$ and
    latency $1$. Note that we construct this graph in polynomial time, as required for the pre-processing phase.

    We maintain the invariant that at the beginning of the $k$th round all
    edges incident to $s$ have time label at least $n-k$; this condition holds
    initially, as there are no edges incident to $s$. Given a pair of vectors
    $u_k$ and $v_k$ for $k\in[n]$, we perform the following insertions:
    \begin{itemize}
        \setlength{\itemsep}{0pt}
        \item for every $i\in[n]$ such that $u_k[i]=1$, update with $\AddEdge((s, x_i), n-k, 1)$
        \item for every $j\in[n]$ such that $v_k[j]=1$, update with $\AddEdge((y_j, t_k), n-k+2, 1)$
    \end{itemize}
    Suppose that after these updates there exists a temporal path $P$ from $s$
    to $t_k$. Then the predecessor of $t_k$ on $P$ is reached by time at most
    $n-k+2$. Since all edges incident to $s$ have time labels at least $n-k$
    and all edges in the graph have latency $1$, this predecessor is reachable
    from $s$ by a path of at most two edges. Hence $P$ must be of the form
    $(s, x_i),\ (x_i, y_j),\ (y_j, t_k)$ for some $i,j\in[n]$, with time
    labels $n-k,\ n-k+1,\ n-k+2$, respectively.

    We now query $\Dist(t_k)$ to check whether $t_k$ is reachable from $s$,
    that is, whether the value returned by the structure is different from
    $\infty$. If this happens, then by the argument above the witnessing path
    must be of the form $(s, x_i),\ (x_i, y_j),\ (y_j, t_k)$ for some
    $i,j\in[n]$ with time labels $n-k,\ n-k+1,\ n-k+2$, respectively. Hence
    $u_k[i]=1$, $M[i][j]=1$, and $v_k[j]=1$, so $u_kMv_k=1$.

    Conversely, if $u_kMv_k=1$, then for some $i,j\in[n]$ there are edges
    $(s, x_i),\ (x_i, y_j),\ (y_j, t_k)$ with time labels
    $n-k,\ n-k+1,\ n-k+2$, respectively. Thus $t_k$ is reachable from $s$.
    The invariant is preserved, since every newly inserted edge incident to $s$
    has time label $n-k$, while all older such edges have larger time labels.
    In each round we perform at most $2n$ updates and one query.

    \subparagraph{Decremental model.} We initialize the structure similarly, with a
    graph $G$ on $3n+1$ vertices
    $s, x_1,x_2,\dots, x_n, y_1, y_2, \dots, y_n, t_1, t_2, \dots, t_n$. For
    every $i, j\in [n]$ such that $M[i][j]=1$, we add temporal edges
    $(x_i, y_j)$ with time labels $2,3,\dots, n+1$ and latency $1$. Moreover, for
    every $i\in[n]$ we add temporal edges $(s, x_i)$ with time labels
    $1,2,\dots, n$ and latency $1$, and for every $j,k\in [n]$ we add temporal
    edges $(y_j, t_k)$ with time label $k+2$ and latency $1$. Note that we construct this graph in polynomial time, as required for the pre-processing phase.

    We maintain the invariant that at the beginning of the $k$th round all
    edges incident to $s$ have time label at least $k$. Then, in each round,
    given a pair of vectors $u_k$ and $v_k$ for $k\in[n]$, we perform the
    following deletions:
    \begin{itemize}
        \setlength{\itemsep}{0pt}
        \item for every $i\in[n]$ such that $u_k[i]=0$, we update with $\RemoveEdge((s, x_i), k, 1)$
        \item for every $j\in[n]$ such that $v_k[j]=0$, we update with $\RemoveEdge((y_j, t_k), k+2, 1)$
    \end{itemize}
    As before, we first describe the structure of any temporal path $P$ from
    $s$ to $t_k$. The predecessor of $t_k$ is reached on $P$ by time at most
    $k+2$. Since all edges incident to $s$ have time labels at least $k$ and
    all edges in the graph have latency $1$, the path $P$ must be of the form
    $(s, x_i),\ (x_i, y_j),\ (y_j, t_k)$ with time labels
    $k,\ k+1,\ k+2$, respectively.

    We query $\Dist(t_k)$ to check whether $t_k$ is reachable from $s$, that
    is, whether the value returned by the structure is different from
    $\infty$. If so, then the temporal path witnessing reachability is of the
    form $(s, x_i),\ (x_i, y_j),\ (y_j, t_k)$. Therefore $u_k[i]=1$,
    $M[i][j]=1$, $v_k[j]=1$, and $u_kMv_k=1$.

    Conversely, if $u_kMv_k=1$, then for some $i,j\in [n]$ there exist edges
    $(s, x_i),\ (x_i, y_j),\ (y_j, t_k)$ with time labels
    $k,\ k+1,\ k+2$, respectively, and therefore $t_k$ is reachable from $s$.
    At the end of each round, we call $\RemoveEdge((s, x_i), k, 1)$ for all
    $i\in[n]$ such that $u_k[i]=1$. Then every edge incident to $s$ has time
    label at least $k+1$, so the invariant is preserved. In each round we
    perform at most $2n$ updates and one query.
\end{proof}

\begin{corollary}\label[corollary]{lem:fully_undirected_lowerbound}
    Assuming the OMv conjecture, for any $\eps > 0$ there is no fully dynamic
    algorithm solving the Single Source Temporal Path Reachability problem on
    undirected graphs with $n$ vertices, with:
    \begin{itemize}
        \setlength{\itemsep}{0pt}
        \item polynomial time preprocessing,
        \item update in $\Oh(n^{1-\eps})$ amortized time, and
        \item query in $\Oh(n^{2-\eps})$ amortized time.
    \end{itemize}
    This lower bound holds even if the algorithm is allowed to output incorrect
    answers with probability at most $1/3$.
\end{corollary}
\begin{proof}
    This is an immediate consequence
    of~\cref{lem:partially_undirected_lowerbound}, since every fully dynamic
    algorithm is, in particular, a partially dynamic algorithm when the update
    sequence is restricted either to insertions only or to deletions only.
\end{proof}

In the constructions above, every temporal edge has latency $1$. The same
constructions also work when every latency is equal to $0$, provided that we
restrict attention to \emph{strict} temporal paths, i.e., temporal paths whose
time labels are strictly increasing.

\begin{lemma}\label{lem:lb}
Assuming the OMv conjecture, for every $\varepsilon>0$ there is no incremental algorithm for
$\Dist(v,t_0)$ (even only deciding whether $\Dist(v,t_0)<\infty$) with polynomial preprocessing, and total 
$O(m^{3/2-\varepsilon})$ running time, on
instances with $m=\Theta(n^2)$ arcs and $\Theta(n)=\Theta(\sqrt{m})$ different departure times from the source. In particular, total update time
$O(m^{3/2-\varepsilon})$ together with query time $O(m^{1-\varepsilon})$ is impossible.
This holds even if all latencies are equal to $1$, and even with error probability $1/3$.
\end{lemma}
\begin{proof}
Let us consider $n=3N+1$ and let us create vertices $s$, $x_1,\dots,x_N$, $y_1,\dots,y_N$, $t_1,\dots,t_N$.
Let $T=N+1$.
 
\subparagraph{Preprocessing.} For every $i,j$ with $M[i][j]=1$ insert the single arc $(x_iy_j,T,1)$.
This gives at most $N^2$ arcs.
 
\subparagraph{Round $k$ (for $k \in \{ 1, \ldots, N \}$).} On input $(u_k,v_k)$:
\begin{itemize}[nosep]
  \item for every $i$ with $u_k[i]=1$ insert $(sx_i,k,1)$;
  \item for every $j$ with $v_k[j]=1$ insert $(y_jt_k,T+1,1)$;
  \item query $\Dist(t_k,k)$ and output $1$ iff the answer is finite.
\end{itemize}
 
\subparagraph{Correctness.} Consider a temporal path $P$ from $s$ to $t_k$ departing at time $\ge k$.
Its first arc leaves $s$ with label $\ge k$; arcs leaving $s$ inserted in rounds $k'<k$ have
label $k'<k$, and later rounds have not happened yet, so the first arc is $(sx_i,k,1)$ with
$u_k[i]=1$. The only arcs entering $t_k$ are $(y_jt_k,T+1,1)$ with $v_k[j]=1$; the only arcs
entering any $y_j$ are matrix arcs $(x_iy_j,T,1)$; the only arcs entering any $x_i$ leave $s$;
vertices $t_{k'}$ have no outgoing arcs. Hence $P=s\,x_i\,y_j\,t_k$ with $u_k[i]=M[i][j]=v_k[j]=1$.
Conversely, for such $i,j$ the path using labels $k,T,T+1$ is temporal, since
$k+1\le N+1=T$ and $T+1\le T+1$; it arrives at time $T+2$. Thus
$\Dist(t_k,k)<\infty \iff u_k^{\top}Mv_k=1$.
 
\subparagraph{Analysis.} There are $m=O(N^2)$ arcs, $O(N^2)$ updates and $N$ queries. By the OMv conjecture, there is no algorithm with polynomial preprocessing and $\mathcal{O}(N^{3-\varepsilon})$ running time solving the OuMv problem. Since an algorithm with $\mathcal{O}(m^{3/2-\varepsilon})$ running time implies an algorithm with $\mathcal{O}(N^{3-2\varepsilon})$ for the OuMv problem, its existence is ruled out if we assume the OMv conjecture. The instance we construct has $m=\Theta(n^2)$ arcs and $\Theta(n)$ different departure times from $s$.
\end{proof}

\section*{Use of generative AI}
We used Chat GPT-6 to proofread the manuscript. All ideas, proofs and the write-up is due to the authors, except of the last lower bound for the bounded time window, which was suggested by Chat GPT during the proofreading of the paper.

\bibliographystyle{plainurl}
\bibliography{biblio}

\end{document}